# A MOSAIC of methods: Improving ortholog detection through integration of algorithmic diversity

M. Cyrus Maher[1] and Ryan D. Hernandez[2,3,4]

[1]Department of Epidemiology and Biostatistics, University of California, San Francisco, 185 Berry Street, Lobby 5, Suite 5700 San Francisco, CA 94107 [2]Department of Bioengineering and Therapeutic Sciences, [3]Institute for Human Genetics, [4]Institute for Quantitative Biosciences (QB3), University of California, San Francisco, 1700 4th Street San Francisco, California 94158

**Abstract:**

**Ortholog detection (OD) is a critical step for comparative genomic analysis of protein-coding sequences. In this paper, we begin with a comprehensive comparison of four popular, methodologically diverse OD methods: MultiParanoid, Blat, Multiz, and OMA. In head-to-head comparisons, these methods are shown to significantly outperform one another 12-30% of the time. This high complementarity motivates the presentation of the first tool for integrating methodologically diverse OD methods. We term this program MOSAIC, or Multiple Orthologous Sequence Analysis and Integration by Cluster optimization. Relative to component and competing methods, we demonstrate that MOSAIC more than *quintuples* the number of alignments for which all species are present, while simultaneously maintaining or improving functional-, phylogenetic-, and sequence identity-based measures of ortholog quality. Further, we demonstrate that this improvement in alignment quality yields 40-280% more confidently aligned sites. Combined, these factors translate to higher estimated levels of overall conservation, while at the same time allowing for the detection of up to 180% more positively selected sites. MOSAIC is available as python package. MOSAIC alignments, source code, and full documentation are available at <http://pythonhosted.org/bio-MOSAIC>.**

**Introduction:**

Orthologs are genes shared between organisms that derive from a common ancestral gene but have diverged from one another through speciation. This is in contrast to paralogs, which arise through gene duplication within a given genome. It is common in comparative genomics and phylogenetics to extract evolutionary information about a particular gene from its alignment with orthologous sequences. To enable this analysis, orthologs must first be inferred, making ortholog detection (OD) an indispensible first step in a variety of phylogenetic inference tasks [1, 2].

In general, existing OD methods can be classified as tree-based, graph-based, or a hybrid of the two (Altenhoff and Dessimoz 2012). Tree-based methods may use reconciliation techniques between gene and species trees or may rely on the gene tree alone. Graph-based methods can employ a variety of metrics to quantify similarity between sequences, including BLAST scores or sequence identity. Information about the conserved gene neighborhood may also be included in this context. Techniques such as Markov clustering may then be applied to create orthologous groups, or one may simply define clusters based on a graph's existing connections (Kuzniar et al. 2008).

Unfortunately, the few benchmarking studies that have sampled broadly from this methodological diversity have provided equivocal results. Although there are general trends in relative effectiveness between methods, performance is highly context-dependent and does not always favor more sophisticated approaches (Hulsen et al. 2006; Chen et al. 2007; Altenhoff and Dessimoz 2009a). This is discouraging from the point of view of identifying a single best OD method, but it also suggests a new and relatively facile avenue for methodological improvement. By harnessing differences between OD methods, a wide variety of algorithms may play complementary roles within a cooperative inference framework.

We begin our analysis with a comprehensive comparison of four popular and methodologically distinct OD methods: 1.) MultiParanoid, a reciprocal-BLAST plus Markov clustering method (Alexeyenko et al. 2006); 2.) TBA, a synteny-based aligner used to produce UCSC's MultiZ alignments (Blanchette et al. 2004); 3.) six-frame translated BLAT, a fast, approximately-scored protein query approach that does not rely on predicted proteomes (Kent 2002); and 4.) OMA, a well-established tree-based method (Altenhoff et al. 2011). Applying these methods to OD in a range of primates and closely related mammals, we demonstrate that methodological performance varies widely by species and appears to depend critically on genome quality.

Next, we characterize the striking performance gains yielded by combining these methods. This is demonstrated using sequence identity, phylogenetic tree concordance, and Hidden Markov Model-based functional agreement. In addition, we show that our approach significantly outperforms metaPhOrs, the existing tree-based approach to OD integration (Pryszcz et al. 2011). Finally, we demonstrate that our improvement in alignment quality translate to higher estimated levels of overall conservation, while at the same time, detecting up to 180% more positively selected sites.

The implementation of this novel approach for the integration of diverse ortholog detection methods is presented as the software tool, MOSAIC, or **M**ultiple **O**rthologous **S**equence **A**nalysis and **I**ntegration by **C**luster optimization. MOSAIC is implemented as a well-documented python package that can be installed using easy_install bio-mosaic from the command-line. MOSAIC alignments, source code, and full documentation are available at http://pythonhosted.org/bio-MOSAIC.

**New Approaches:**

*OD integration as cluster optimization*

MOSAIC provides a highly flexible, graph-based framework for integrating diverse OD methods. Proposal orthologs are conceptualized as nodes in a graph, connected with edges weighted according to the pairwise similarity between sequences. The task of OD integration is then to choose proposal orthologs for each sequence such that a chosen measure of intra-cluster similarity is optimized.

*MOSAIC optimizes (weighted) pairwise similarities*

To begin, MOSAIC calculates pairwise similarities between all orthologs from different species. Percent identity- and blast-based similarity metrics are provided by default, but user-defined similarity metrics are also accepted. These similarity scores define edge weights, which are used to construct a graph such as the one presented at the top of Box 1. Once this full graph is constructed and quality filtered, MOSAIC then chooses at most one proposal ortholog from each species so that the overall pairwise similarity between accepted sequences is optimized.

To accommodate user priorities, pairwise similarities can be weighted such that sequences from different species contribute unequally to the total similarity score. For uniform weights, this is equivalent to maximizing the average pairwise similarity. In the case where only similarity to a reference sequence is of interest, this reduces to simply taking the sequence for each species that is most similar to the reference.

*Optimization is carried out using cyclic coordinate descent*

For *m* OD methods and *s* species, there are up to $m^s$ possible integrated alignments. In the case analyzed in this paper, *m=4* and *s=10*. This translates to over a million possible integrated alignments for each of the ~25,000 reference sequences considered. It is clear to see from this example that an exhaustive optimization becomes quickly infeasible. Therefore, MOSAIC choses optimal clusters using cyclic coordinate descent (CCD), an efficient non-derivative optimization algorithm (Bertsekas 1999).

In Box 1, we illustrate the way CCD functions in the context of MOSAIC. After building the full graph that includes all orthologous sequences, random orthologs from each species are chosen as the current best. MOSAIC then loops through the species of interest in a random order. For each species, MOSAIC choses the sequence that optimizes cluster tightness given the current best sequences for all other species. This process is repeated until no further improvements can be made to cluster tightness. Finally, since CCD is prone to finding local rather than global optima, this entire process is repeated multiple times with random starting points and sampling paths.

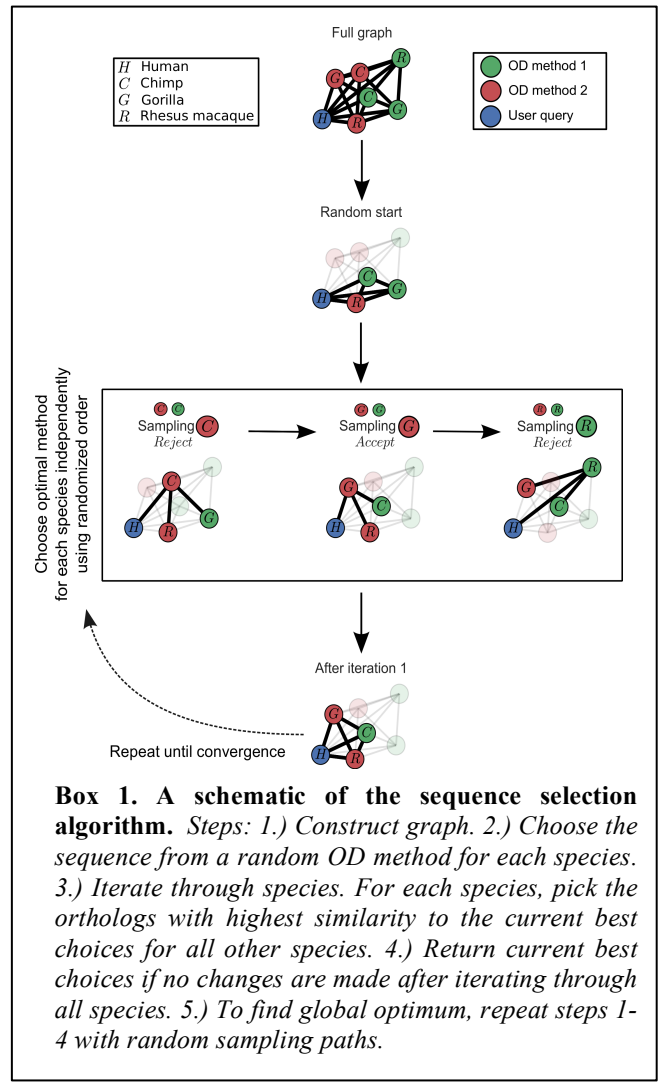

**Box 1. A schematic of the sequence selection algorithm.** *Steps: 1.) Construct graph. 2.) Choose the sequence from a random OD method for each species. 3.) Iterate through species. For each species, pick the orthologs with highest similarity to the current best choices for all other species. 4.) Return current best choices if no changes are made after iterating through all species. 5.) To find global optimum, repeat steps 1-4 with random sampling paths.*

**Results and Discussion:**

*Ortholog detection methods frequently outperform one another*

To motivate OD integration, we will begin with a comprehensive comparison of four popular, methodologically diverse OD methods. In figure 1, we show the head-to-head performances of these different methods for a range of primates and closely related mammals. Performance is assessed on alignments generated between all human consensus coding sequences (CCDS) (Pruitt et al. 2009) and their corresponding orthologs. More specifically, for each possible orthologous sequence, we examine the proportion of orthologs from all species for which the level of sequence identity to human is at least five percentage points higher for one particular method versus another, otherwise we consider the two methods to be tied. By this metric, we observe that one method significantly outperforms another 10% to 30% of the time. Importantly, no method uniformly outperforms all others, underlining the complementarity of the chosen methods.

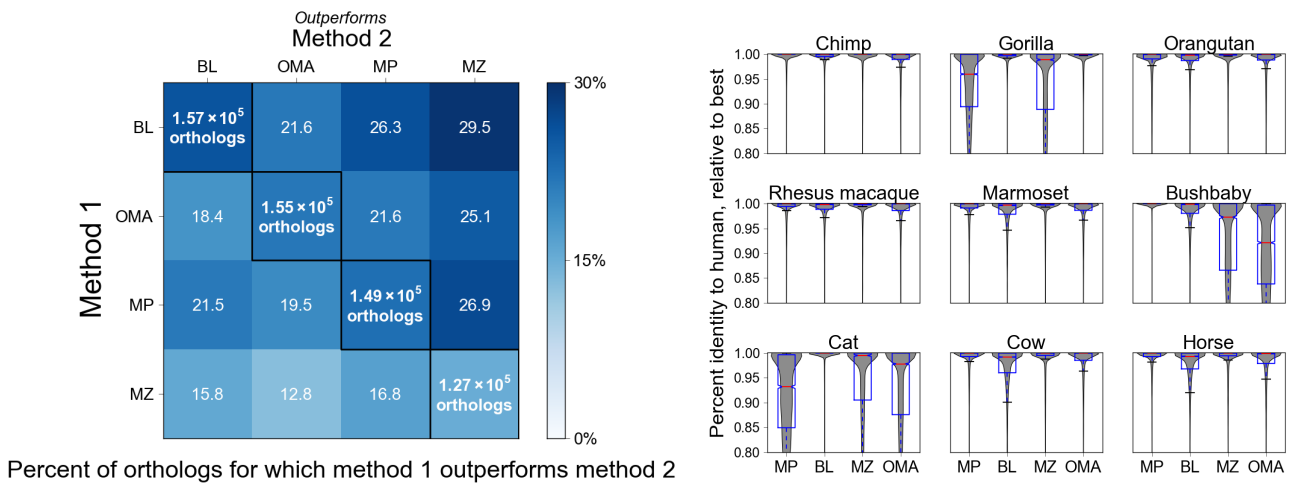

*Figure 1. Comparison of sequence identity levels between methods* A.) Heat map of the percent of orthologs for which MultiParanoid (MP), OMA (OMA), BLAT (BL) and MultiZ (MZ) outperform one another. Performance is based on percent identity of each method's orthologs to the human sequence. One method is considered to outperform another method if it improves percent identity by at least five percentage points. Text in diagonal cells shows the number of orthologs identified by each method, colored by the percent of transcripts at which a given method outperforms all the others. B.) Distributions of percent identity relative to the highest scoring ortholog, stratified by species.

We next evaluated percent identity to human for each ortholog proposed by each method relative to the highest scoring ortholog from all methods. figure 2A demonstrates that relative performance is species-specific. In particular, we note that the performance disparities across methods are much more pronounced for gorilla, bushbaby, and cat, both in terms of the number and quality of obtained orthologs.

Examining each OD method in detail yields some hypotheses about the origin of these differences in performance. Errors in proteome prediction, both in terms of false-positives and false-negatives, are likely to have large effects on both MultiParanoid and OMA. Meanwhile, spurious syntenic information is expected to compromise the integrity of ortholog predictions produced by MultiZ. Finally, the lack of an assembled genome for bushbaby may negatively impact the quality of the one-way BLAT approach due to the segmentation of exon sets across multiple unordered scaffolds.

*Combining multiple sequence alignments with MOSAIC*

It is well-known in theory (Wolpert and Macready 1997) and in practice (van der Laan and Gruber 2010) that the comparative performance of competing statistical inference algorithms often varies by context. Rather than search for a single best algorithm, researchers have sought to integrate a variety of methods in order to reap the benefits of methodological complementarity (van der Laan et al. 2007; Rokach 2009; Chandrasekaran and Jordan 2013). As might be expected, the gains yielded by this approach generally scale with the quality of the individual methods integrated, the number of methods included, and, importantly, the diversity of the comprised algorithms (Kuncheva and Whitaker 2003).

Having observed the complementarity between the OD methods presented above, we sought to develop a structure for the automatic integration of methodologically diverse OD methods such as those described above. We term this framework MOSAIC, or **M**ultiple **O**rthologous **S**equence **A**nalysis and **I**ntegration by **C**luster optimization. MOSAIC allows for the flexible integration of diverse OD methods through the application of standard or user-defined metrics of sequence divergence and ortholog cluster quality. Using specified divergence metrics, clusters of proposed orthologs are built. These orthologs are then adopted or rejected in order to optimize cluster completeness and quality (e.g., distance to a reference sequence or average pairwise distance).

For the examples presented here, we consider a protein set with relatively low levels of evolutionary divergence, and so choose percent identity as our metric for sequence divergence. However, for more distantly related species, the application of scoring matrices (Dayhoff et al. 1978; Henikoff 1992) or Hidden Markov Models (Ebersberger et al. 2009) may be preferable for measuring divergence. For each human sequence, each method may propose an ortholog from each species. Corresponding putative orthologs are then evaluated according to the percent of sites in the human protein sequence that are identical to it (indels and substitutions are penalized equivalently per base, though affine scoring could be accommodated). The best-scoring ortholog among all methods is then chosen for each species.

*Combining methods increases the number of included sequences*

To assess the efficacy of MOSAIC, we first examined the total number of species included in alignments to human CCDS sequences. For MOSAIC and each OD method, we observe the number of alignments to human CCDS as a function of the maximum number of missing species allowed. Strikingly, the integration of methods more than quintuples the number of alignments for which all species are present (fig. 2B). As expected, the gains afforded by MOSAIC are species-specific and increase as a function of the number of methods that are included (fig. 2A). Using MultiZ as a baseline, we observe once again that the largest improvements are seen for gorilla, bushbaby, and cat. Importantly, orthologs for each of these three species are rescued by different methods (OMA for gorilla, MultiParanoid for bushbaby, and BLAT for cat), further demonstrating the power of integrating diverse OD methods.

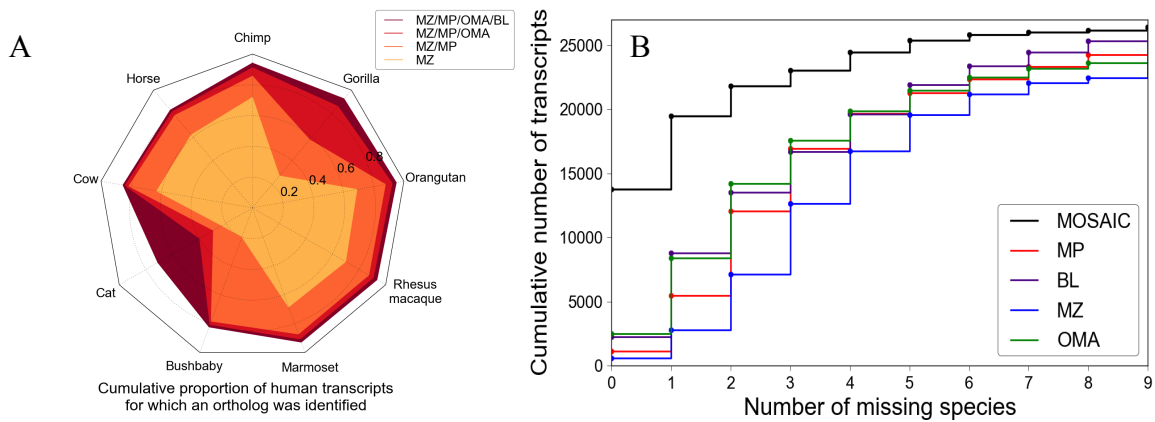

*Figure 2. OD power and the effect of pooling methods* A.) The cumulative proportion of human transcripts for which an ortholog was detected, stratified by species. Envelopes illustrate results from pooling an increasing number of methods. B.) The cumulative number of human transcripts as a function of the maximum number of missing species allowed.

*MOSAIC adds new sequences, maintains or increases average levels of sequence identity*

Figure 3 demonstrates that, for each species, MOSAIC retrieves a much larger number of sequences than any method alone, while maintaining levels of percent identity comparable to those of the best performing method. It should be noted here that in our current examples, MOSAIC is designed to optimize the metric of sequence identity to human. Indeed, for a given putative ortholog, MOSAIC is guaranteed to improve or maintain percent identity compared to its constituent methods. Counterintuitively, this provides no assurance that MOSAIC will provide gains in *average* levels of percent identity. For example, average levels of percent identity could decrease if MOSAIC ensures the inclusion of a greater number of species by pulling in poorly scoring sequences that were initially filtered out by the majority of component methods. We see however that this is not the case.

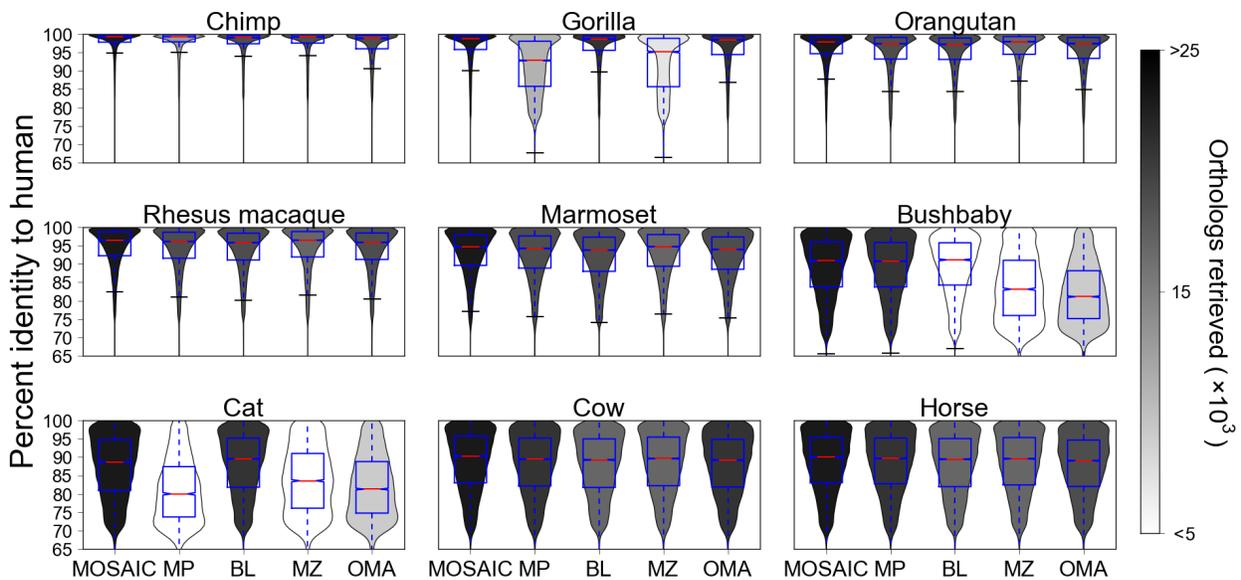

*Figure 3. The effect of method integration on sequence identity.* *A comparison of the overall distributions of percent identity to human for MOSAIC and its component methods. As in figure 1B, smoothed distributions underlying the boxplots are shaded according to the number of human transcripts for which an ortholog was proposed. White denotes 5000 sequences or less. Darker shades signify increasingly larger numbers of detected orthologs.*

*Integrating methods leads to higher levels of phylogenetic and functional concordance*

To further examine the effect of MOSAIC on alignment quality, we compared phylogenetic and functional concordance across methods. Phylogenetic concordance was ascertained by calculating the normalized, unweighted Robinson-Foulds (RF) distance (Robinson and Foulds 1981) between gene trees and the established species tree (Altenhoff and Dessimoz 2009b). This metric is equal to the sum of the number of splits in one tree that are not present in the other, scaled by the total number of splits present across the two trees. Accordingly, larger RF distances correspond to worse agreement between gene and species trees. On a gene-by-gene basis, this metric should be interpreted with caution, since post-speciation admixture and incomplete lineage sorting can lead to true discordance between the species tree and the phylogenetic history of a particular gene (Maddison and Knowles 2006). However, at the level of the genome, higher concordance between gene trees and the known speciation process strongly suggests a relative improvement in OD.

Figure 4A presents the cumulative proportion of alignments included as a function of the maximum allowable RF distance. Multiz is seen to perform the best of any individual method, likely due to its utilization of syntenic information. Surprisingly, the tree-based OD method, OMA, is seen to be the worst performing method according to this tree-based metric. Combining all methods using MOSAIC leads to a strong enrichment of highly concordant gene trees, while providing performance that is competitive with all component methods at more permissive RF distance cutoffs.

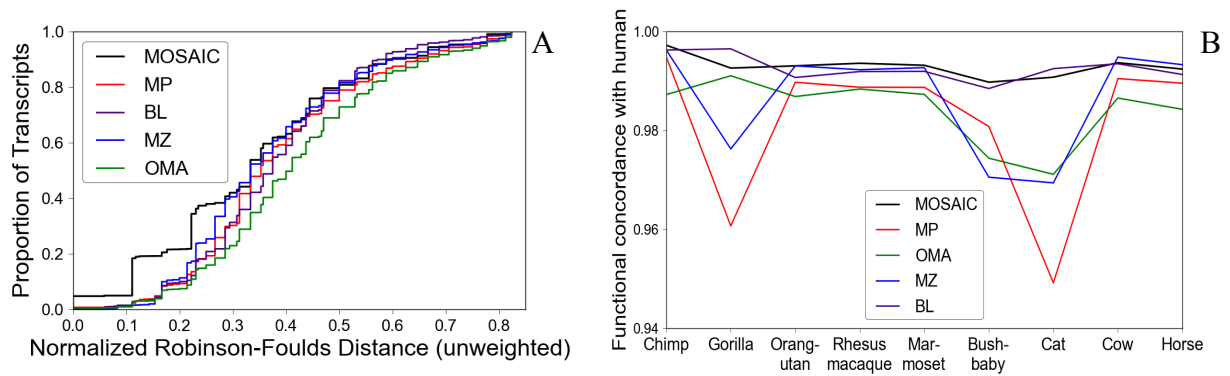

*Figure 4. The effect of method integration on tree-based and functional concordance.* *A.) The cumulative proportion of human transcripts as a function of the maximum allowable Robinson-Foulds distance between the*

*gene tree and the species tree. B.) The rate of concordance between functional annotations for proposal orthologs and human transcripts.*

In addition, we used profile HMMs from the Protein Families Database A (PfamA) (Punta et al. 2012) and HMMER3 (Eddy 2011) to ascertain functional concordance between proposed orthologs and the human CCDS of interest. PfamA builds HMMs via curated alignments of small numbers of representative members from each protein family. Using HMMER3, we queried protein sequences against all PfamA protein family profiles, annotating each protein according to its top protein family hit. This allowed for an ascertainment of functional concordance that is vastly more comprehensive than relying on gene-by-gene annotation across species, while retaining many of the advantages of manual curation. This assessment reveals that, for the set of orthologous sequences proposed by all methods, MOSAIC provides levels of functional concordance that are comparable to the best performing method and considerably better than most methods for gorilla, bushbaby, and cat (fig. 4B).

*MOSAIC outperforms tree-based OD integration, even by tree-based metrics*

We have shown that MOSAIC provides a large increase in the number of detected orthologs relative to its component methods, while simultaneously maintaining or improving functional-, phylogenetic-, and sequence identity-based measures of ortholog quality. Next, we sought to compare this method of OD integration to the only alternative of which we are aware: metaPhOrs (Pryszcz et al. 2011). Using an approach based on tree overlap, metaPhOrs integrates ortholog predictions using phylogenetic trees from seven databases: PhylomeDB, Ensembl, TreeFam, EggNOG, OrthoMCL, COG, and Fungal Orthogroups.

We compared MOSAIC and metaPhOrs based on the number of retrieved orthologs, average differences in sequence identity, and comparative levels of functional and phylogenetic concordance. We observe that MOSAIC provides large increases in the number of retrieved orthologs, while providing slight improvements in sequence identity for those cases where proposal orthologs are available from both methods (fig. 5). For the cases where MOSAIC predicted an ortholog but metaPhOrs did not, we examined the level of sequence identity in these sequences compared to the species-specific average returned by metaPhOrs.

We find that these additional sequences display levels of sequence identity comparable to those provided by metaPhOrs. Finally, we observe that MOSAIC yields a slight increase in functional concordance, as well as a 40% increase in tree concordance, measured as the area under the curve below an RF distance of 0.5. A 0.5 threshold was chosen because there is little differentiation between methods after this point.

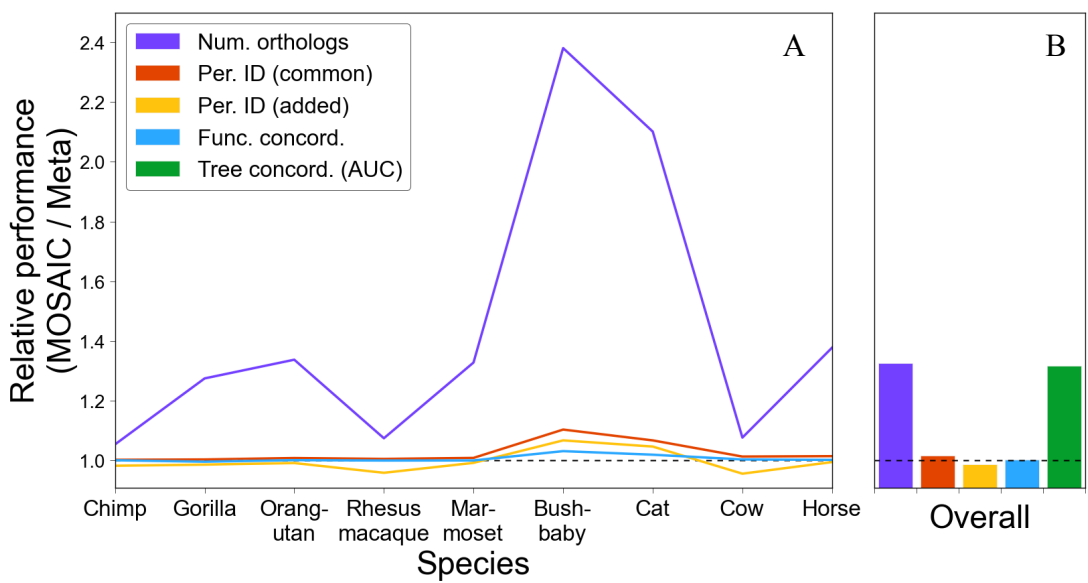

*Figure 5. A comparison between MOSAIC and metaPhOrs.* *The relative performance between MOSAIC and metaPhOrs according to five metrics: 1.) the number of orthologs detected (purple); 2.) the percent identity to human for orthologs present in both (red); 3.) the percent identity to human for orthologs unique to MOSAIC compared to metaPhOrs species-specific average (yellow); 4.) rate of functional concordance between proposal orthologs and human transcripts (blue); and 5.) concordance between gene and species trees, as measured by a normalized, unweighted Robinson-Foulds distance (green). A.) The breakdown of relative performance by species. B.) Relative performance averaged across species. Scale is matched to panel A. Note that tree concordance is only included in panel B because it is calculated based upon full sequence alignments.*

*Increased ortholog quality leads to more conservation and more positively selected sites*

Having demonstrated an increase in ortholog quality using tree-, function-, and similarity-based measures of quality, we next sought to assess the influence of increased alignment quality on estimated levels of selection. To assess gene-level conservation, we applied Phylogenetic Analysis by Maximum Likelihood (PAML) (Yang 2007) with automated likelihood-based model selection. To ascertain site-level positive selection, we employed Sitewise Likelihood Ratio (SLR), a method shown to have a higher power and a lower false positive rate than PAML's popular Bayes Empirical Bayes (BEB) method (Massingham and Goldman 2005).

Since varying numbers of sequences can sway evolutionary estimates in unpredictable ways due to, e.g. inhomogeneous levels of selection across organisms, we assessed the performance of MOSAIC relative to each method by matching the species present in each alignment. We refer to this approach as MOSAIC$_{matched}$. In the case of both PAML and SLR, synonymous substitution rates in coding DNA are used as a background against which to test for changes in rates of non-synonymous substitution. Since the metaPhOrs database provides only protein sequences for its alignments, no comparison with this method was possible given the available data.

In figure 6A, we see that MOSAIC leads to higher gene-level conservation (lower dN/dS) compared to every method except Blat, for which the difference was not statistically significant. Despite higher levels of conservation, however, MOSAIC was able to detect between ~30 and 180% more positively selected sites than any of its component methods. Rather than an increase in the fraction of sites detected as being positively selected, most of this increase in power was due to the fact that more sites were aligned to high confidence and therefore included in the analysis. This step of filtering for alignment quality is important because site-wise estimates of positive selection are highly sensitive to short poorly aligned regions (Jordan and Goldman 2012).

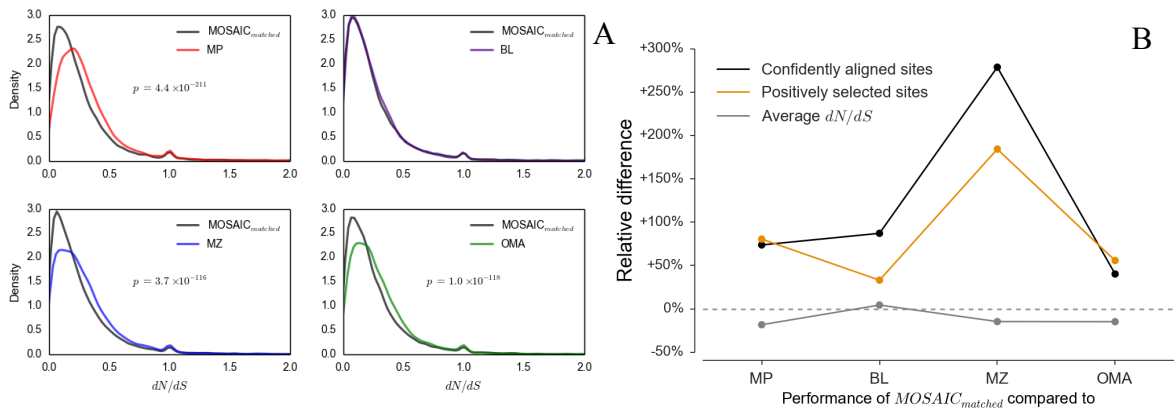

*Figure 6. A comparison of evolutionary estimates.* To facilitate an unbiased comparison, MOSAIC alignments were produced to match the species included by each component method (referred to as MOSAIC$_{matched}$). A.) The distribution of gene-level conservation (measured by dN/dS) for each component method versus MOSAIC$_{matched}$ B.) The relative difference of MOSAIC$_{matched}$ versus each component method for: 1.) the number of positively

*selected sites, 2.) the number of confidently aligned sites, and for reference, 3.) the average level of conservation across all alignments.*

*Conclusions*

In this paper we have introduced a novel algorithm, MOSAIC, which is capable of integrating an arbitrary number of methodologically diverse ortholog detection methods. We have demonstrated that MOSAIC provides large increases in power relative to its component methods, while simultaneously maintaining or improving functional-, phylogenetic-, and sequence identity-based measures of ortholog quality. Furthermore, MOSAIC is the best performing OD integration method and the only one that can easily produce alignments of coding DNA. These integrated alignments often include more species than any component method. Further, given the same number of species, MOSAIC alignments include more columns aligned with high confidence. This translates to higher levels of estimated conservation, and simultaneously, a greatly increased number of positively selected sites detected.

In summary, MOSAIC provides the unique flexibility to incorporate any OD method, thereby increasing methodological diversity, improving OD performance, and allowing researchers to take advantage of methodological gains in a variety of areas of OD research. MOSAIC is a python package that can be installed using easy_install bio-mosaic from the command-line. MOSAIC alignments, source code, and full documentation are available at http://pythonhosted.org/bio-MOSAIC.

**Materials and methods:**

*Retrieval of orthologs*

For each human consensus coding sequence (version GRCh37.p9), we sought to retrieve orthologs for chimp, gorilla, orangutan, rhesus macaque, marmoset, bushbaby, cat, cow, and horse. For MultiParanoid, MultiZ, and Blat, genomic data was retrieved for the following genome builds:

| Genome | Version | Release |
|---|---|---|
| Chimp | panTro4 | Feb-11 |
| Gorilla | gorGor3.1 | May-10 |
| Orangutan | ponAbe2 | Jul-11 |
| Rhesus macaque | rheMac3 | Oct-10 |
| Marmoset | calJac3 | Mar-09 |
| Bushbaby | otoGar3 | May-11 |
| Cat | felCat5 | Sep-11 |
| Cow | bosTau7 | Oct-11 |
| Horse | equCab2 | Sep-07 |

For MultiParanoid (Alexeyenko et al. 2006), an all-versus-all blast search was run using the following command structure:

blastp -db $blastdatabase -query [query file] -out [output file] -evalue .01 -num_threads [number of threads] -outfmt 6 -db_soft_mask 21 -word_size 3 -use_sw_tback

From this output, ortholog predictions were produced using the standard MultiParanoid protocol.

For BLAT (Kent 2002), genomes for each species of interest were downloaded from the NCBI Entrez Genome database (McEntyre and Ostell 2002). Queries were conducted using the following command structure:

blat -q=prot -t=dnax -minIdentity=70 –extendThroughN [genome file] [query file] [output file]

In the case of MultiZ (Blanchette et al. 2004), CCDS orthologs were downloaded directly from the UCSC genome browser (Kent et al. 2002). For OMA (Altenhoff et al. 2011), ortholog predictions were downloaded from omabrowser.org (December 2012 release). For genes with more than one CCDS, orthologs were mapped to each analyzed transcript. Finally, ortholog predictions from metaPhOrs (Pryszcz et al. 2011) were retrieved from release v201009 (June 2012).

To remove possibly spurious orthologs, proposals from each method were then filtered according to a species-specific sequence identity cutoff, as described below.

*Filtering and integration of orthologs*

For each proposed ortholog for a given CCDS, the CCDS and the orthologous sequence under consideration were globally realigned using the program stretcher from the EMBOSS toolkit (Rice et al. 2000). Percent identity was then calculated as the percent of sites in the human sequence that were identical in the orthologous sequence. For example, the hypothetical sequence below would be scored as 71% identical (5/7), since there are 2 mismatches between the seven sites present in the human sequence and the character to which those sites are aligned in the chimp sequence (sites where the human sequence has been deleted or the outgroup has an insertion are ignored):

```
Human   A W V A - T F D

Chimp   - W V R Y T F D
```

All orthologs with percent identity below a critical threshold were removed from all subsequent analyses. This cutoff was chosen considering the known level of genome-wide divergence between human and the species of interest, as well as the overall distributions of percent identity between putative orthologs in the two species. These cutoffs were as follows: chimp: 82%, gorilla: 77%, orangutan: 75%, rhesus macaque: 73%. A cutoff of 70% was employed for marmoset, bushbaby, cat, cow, and horse.

After filtering, the orthologs with the highest percent identity from each species were accepted into the integrated orthologous cluster. These sequences were then aligned using MSAprobs (Liu et al. 2010).

*Sequence alignment*

All sequences were aligned using MSAprobs, a multithreaded aligner with better performance benchmarks than many top aligners, including ClustalW, MAFFT, MUSCLE, ProbCons, and Probalign (Liu et al. 2010). Importantly, MSAprobs has the further advantage of providing, for each column of an alignment, dependable estimates of the confidence of the alignment at the site.

*Quality assessment*

*Sequence identity*

MOSAIC optimizes pairwise sequence similarity. In this example, sequence identity is used as the similarity measure, and pairwise similarities are weighted such that only concordance with the human reference sequence is considered. To achieve greater separation between metrics used for optimization and assessment, comparisons of sequence identity were performed in the context of the full multiple sequence alignments (MSAs). We believe this choice is sensible because it is the quality of the MSA that is of primary importance to many downstream phylogenetic inference tasks. In addition, this approach allows us to indirectly incorporate information about intra-cluster similarity, since due the increased number of tradeoffs involved, MSAs between divergent sequences are expected to exhibit a larger degradation in performance relative to pairwise alignments.

*Tree concordance*

For each MSA, gene trees were built using RAxML (Stamatakis and Alachiotis 2010). An unweighted Robinson-Foulds (RF) distance (Robinson and Foulds 1981) was then calculated between each gene tree and the known species tree using the python module dendropy (Sukumaran and Holder 2010). Briefly, the unweighted RF distance counts the number of operations required to transform one tree into the other. This quantity is equal to the total number of splits that are present in one tree but not the other. To normalize for variations in tree size, we then divided this distance by the sum of the total number of splits in the gene and species trees (Yu et al. 2011).

*Functional concordance*

Profile HMMs were downloaded from the PfamA protein families database (Punta et al. 2012). Each sequence was then annotated using the top scoring function retrieved by querying that sequence against the database of all PfamA protein family HMMs. This search was conducted using HMMER3 (Eddy 2011). Functional concordance was then measured as a binary quantity, corresponding to whether or not a putative orthologous sequence had the same inferred function as its cognate human sequence.

*Evolutionary analysis*

*Gene-level conservation*

Alignments were analyzed using Phylogenetic Analysis by Maximum Likelihood (PAML) (Yang 2007). For each alignment three models were fit: 1.) a neutral model where dN/dS is fixed at one, 2.) a conservation model where dN/dS is less than or equal to one, and 3.) a positive selection model where some fraction of the sequence is fit under the conservation model, while dN/dS is estimated freely for the remainder of the sequence. Since evolutionary models are not in general nested, we performed model selection via the popular Akaike Information Criterion (AIC), a method that penalizes a model's fit by its number of included variables (Akaike 1973) and is asymptotically equivalent to maximizing the model's predictive performance on unseen data (Stone 1977).

Despite rigorous model selection procedures, in rare cases PAML may estimate very high levels of selection over a tiny proportion of a given sequence (even a single site), leading to greatly inflated average levels of dN/dS. To reduce the influence of outlying estimates of selection, all dN/dS values greater than 3 were excluded for the analysis. For all methods, this corresponded to less than .05% of all sequences.

*Site-level positive selection*

The program Sitewise Likelihood Ratio (SLR) (Massingham and Goldman 2005) was used to estimate the number of positively selected sites in each sequence. To eliminate false positives due to poorly aligned sites, we filtered out all sites estimated by MSAprobs to be aligned to less than 95% confidence. All included positively selected sites estimated at 95% confidence or greater by SLR were included in the subsequence comparison.

*Plotting*

All data were plotted using the python module matplotlib (Hunter 2007).

**List of abbreviations:**

Ortholog detection (OD), multiple sequence alignment (MSA), hidden markov model (HMM), Robinson-Foulds (RF), MultiParanoid (MP), MultiZ (MZ), BLAT (BL), OMA (OM), cyclic coordinate descent (CCD), Phylogenetic Analysis by Maximum Likelihood (PAML), Sitewise Likelihood Ratio (SLR), Bayes Empirical Bayes (BEB)

**Competing interests:**

The authors have no competing interests to declare.

**Authors' contributions:**

**MCM and RDH conceived the study. MCM wrote the software and gathered and analyzed the data. MCM and RDH drafted the paper. Both authors read and approved the final manuscript.**

**Acknowledgements:**

The authors would like to thank Raul Torres, Lawrence Uricchio, Nicolas Strauli, and Zachary Szpiech for their feedback regarding this manuscript. This work was partially supported by the National Institutes of Health (grant numbers P60MD006902, UL1RR024131, 1R21HG007233, 1R21CA178706, and 1R01HL117004 to R.D.H.). M.C.M. was supported by the Epidemiology and Translational Science program at the University of California, San Francisco, a National Institutes of Health F31 Predoctoral Fellowship (grant number 1 F31 CA180609-01), and a University of California, San Francisco Lloyd M. Kozloff Fellowship.

**References:**

Akaike H. 1973. Information theory and an extension of the maximum likelihood principle. In: Czaki F, Petrov BN, editors. 2nd International Symposium on Information Theory. Budapest: Akademiai Kiado. p. 267–281.

Alexeyenko A, Tamas I, Liu G, Sonnhammer ELL. 2006. Automatic clustering of orthologs and inparalogs shared by multiple proteomes. Bioinformatics [Internet] 22:e9–15. Available from: http://www.ncbi.nlm.nih.gov/pubmed/16873526

Altenhoff AM, Dessimoz C. 2009a. Phylogenetic and functional assessment of orthologs inference projects and methods. PLoS Comput. Biol. [Internet] 5:e1000262. Available from: http://www.pubmedcentral.nih.gov/articlerender.fcgi?artid=2612752&tool=pmcentrez&rendertype=abstract

Altenhoff AM, Dessimoz C. 2009b. Phylogenetic and functional assessment of orthologs inference projects and methods.Eisen JA, editor. PLoS Comput. Biol. [Internet] 5:e1000262. Available from: http://www.pubmedcentral.nih.gov/articlerender.fcgi?artid=2612752&tool=pmcentrez&rendertype=abstract

Altenhoff AM, Dessimoz C. 2012. Inferring Orthology. In: Anisimova M, editor. Evolutionary Genomics. Vol. 855. Methods in Molecular Biology. Totowa, NJ: Humana Press. Available from: http://www.springerlink.com/index/10.1007/978-1-61779-582-4

Altenhoff AM, Schneider A, Gonnet GH, Dessimoz C. 2011. OMA 2011: orthology inference among 1000 complete genomes. Nucleic Acids Res. [Internet] 39:D289–94. Available from: http://www.pubmedcentral.nih.gov/articlerender.fcgi?artid=3013747&tool=pmcentrez&rendertype=abstract

Bertsekas D. 1999. Nonlinear Programming [Hardcover]. Athena Scientific; 2nd edition Available from: http://www.amazon.com/Nonlinear-Programming-Dimitri-P-Bertsekas/dp/1886529000

Blanchette M, Kent WJ, Riemer C, et al. 2004. Aligning multiple genomic sequences with the threaded blockset aligner. Genome Res. [Internet] 14:708–715. Available from: http://www.pubmedcentral.nih.gov/articlerender.fcgi?artid=383317&tool=pmcentrez&rendertype=abstract

Chandrasekaran V, Jordan MI. 2013. Computational and statistical tradeoffs via convex relaxation. Proc. Natl. Acad. Sci. U. S. A. [Internet] 110:E1181–90. Available from: http://www.pubmedcentral.nih.gov/articlerender.fcgi?artid=3612621&tool=pmcentrez&rendertype=abstract

Chen F, Mackey AJ, Vermunt JK, Roos DS. 2007. Assessing performance of orthology detection strategies applied to eukaryotic genomes. PLoS One [Internet] 2:e383. Available from: http://www.pubmedcentral.nih.gov/articlerender.fcgi?artid=1849888&tool=pmcentrez&rendertype=abstract

Ciccarelli FD, Doerks T, von Mering C, Creevey CJ, Snel B, Bork P. 2006. Toward automatic reconstruction of a highly resolved tree of life. Science [Internet] 311:1283–1287. Available from: http://www.ncbi.nlm.nih.gov/pubmed/16513982

Dayhoff MO, Schwartz RM, Orcutt BC. 1978. A model of evolutionary change in proteins. In: Dayhoff MO, editor. Atlas of protein sequence and structure. Nature Biomedical Research. p. 345–358.

Ebersberger I, Strauss S, von Haeseler A. 2009. HaMStR: profile hidden markov model based search for orthologs in ESTs. BMC Evol. Biol. [Internet] 9:157. Available from: http://www.biomedcentral.com/1471-2148/9/157

Eddy SR. 2011. Accelerated Profile HMM Searches.Pearson WR, editor. PLoS Comput. Biol. [Internet] 7:e1002195. Available from: http://dx.plos.org/10.1371/journal.pcbi.1002195

Henikoff S. 1992. Amino Acid Substitution Matrices from Protein Blocks. Proc. Natl. Acad. Sci. [Internet] 89:10915–10919. Available from: http://www.pnas.org/content/89/22/10915

Hulsen T, Huynen MA, de Vlieg J, Groenen PMA. 2006. Benchmarking ortholog identification methods using functional genomics data. Genome Biol. [Internet] 7:R31. Available from: http://www.pubmedcentral.nih.gov/articlerender.fcgi?artid=1557999&tool=pmcentrez&rendertype=abstract

Hunter JD. 2007. Matplotlib: A 2D graphics environment. Comput. Sci. Eng. 9:90–95.

Jordan G, Goldman N. 2012. The effects of alignment error and alignment filtering on the sitewise detection of positive selection. Mol. Biol. Evol. [Internet] 29:1125–1139. Available from: http://www.ncbi.nlm.nih.gov/pubmed/22049066

Kent WJ, Sugnet CW, Furey TS, Roskin KM, Pringle TH, Zahler AM, Haussler a. D. 2002. The Human Genome Browser at UCSC. Genome Res. [Internet] 12:996–1006. Available from: http://genome.cshlp.org/content/12/6/996.abstract

Kent WJ. 2002. BLAT--the BLAST-like alignment tool. Genome Res. [Internet] 12:656–664. Available from: http://www.pubmedcentral.nih.gov/articlerender.fcgi?artid=187518&tool=pmcentrez&rendertype=abstract

Kuncheva LI, Whitaker CJ. 2003. Measures of Diversity in Classifier Ensembles and Their Relationship with the Ensemble Accuracy. Mach. Learn. 51:181–207.

Kuzniar A, van Ham RCHJ, Pongor S, Leunissen J a M. 2008. The quest for orthologs: finding the corresponding gene across genomes. Trends Genet. [Internet] 24:539–551. Available from: http://www.ncbi.nlm.nih.gov/pubmed/18819722

Van der Laan MJ, Gruber S. 2010. Collaborative double robust targeted maximum likelihood estimation. Int. J. Biostat. [Internet] 6:Article 17. Available from: http://www.pubmedcentral.nih.gov/articlerender.fcgi?artid=2898626&tool=pmcentrez&rendertype=abstract

Van der Laan MJ, Polley EC, Hubbard AE. 2007. Super learner. Stat. Appl. Genet. Mol. Biol. [Internet] 6:Article25. Available from: http://www.ncbi.nlm.nih.gov/pubmed/17910531

Liu Y, Schmidt B, Maskell DL. 2010. MSAProbs: multiple sequence alignment based on pair hidden Markov models and partition function posterior probabilities. Bioinformatics [Internet] 26:1958–1964. Available from: http://www.ncbi.nlm.nih.gov/pubmed/20576627

Maddison WP, Knowles LL. 2006. Inferring phylogeny despite incomplete lineage sorting. Syst. Biol. [Internet] 55:21–30. Available from: http://sysbio.oxfordjournals.org/content/55/1/21.abstract

Massingham T, Goldman N. 2005. Detecting amino acid sites under positive selection and purifying selection. Genetics [Internet] 169:1753–1762. Available from: http://www.pubmedcentral.nih.gov/articlerender.fcgi?artid=1449526&tool=pmcentrez&rendertype=abstract

McEntyre .J, Ostell J eds. 2002. The NCBI Handbook. Bethesda, MD: National Center for Biotechnology Information

Pruitt KD, Harrow J, Harte RA, et al. 2009. The consensus coding sequence (CCDS) project: Identifying a common protein-coding gene set for the human and mouse genomes. Genome Res. [Internet] 19:1316–1323. Available from: http://www.pubmedcentral.nih.gov/articlerender.fcgi?artid=2704439&tool=pmcentrez&rendertype=abstract

Pryszcz LP, Huerta-Cepas J, Gabaldón T. 2011. MetaPhOrs: orthology and paralogy predictions from multiple phylogenetic evidence using a consistency-based confidence score. Nucleic Acids Res. [Internet] 39:e32. Available from:

http://nar.oxfordjournals.org/content/39/5/e32.abstract?ijkey=20f551a6b4211ea3241b493f0617a34093ce0154&keytype2=tf_ipsecsha

Punta M, Coggill PC, Eberhardt RY, et al. 2012. The Pfam protein families database. Nucleic Acids Res. [Internet] 40:D290–301. Available from: http://nar.oxfordjournals.org/content/40/D1/D290.full

Rice P, Longden I, Bleasby A. 2000. EMBOSS: the European Molecular Biology Open Software Suite. Trends Genet. [Internet] 16:276–277. Available from: http://www.ncbi.nlm.nih.gov/pubmed/10827456

Robinson DF, Foulds LR. 1981. Comparison of phylogenetic trees. Math. Biosci. [Internet] 53:131–147. Available from: http://dx.doi.org/10.1016/0025-5564(81)90043-2

Rokach L. 2009. Ensemble-based classifiers. Artif. Intell. Rev. [Internet] 33:1–39. Available from: http://link.springer.com/10.1007/s10462-009-9124-7

Stamatakis A, Alachiotis N. 2010. Time and memory efficient likelihood-based tree searches on phylogenomic alignments with missing data. Bioinformatics [Internet] 26:i132–9. Available from: http://www.pubmedcentral.nih.gov/articlerender.fcgi?artid=2881390&tool=pmcentrez&rendertype=abstract

Stone M. 1977. An Asymptotic Equivalence of Choice of Model by Cross-validation and Akaike's Criterion. J. R. Stat. Soc. [Internet]. Available from: http://www.jstor.org/discover/10.2307/2984877?uid=3739560&uid=2134&uid=2&uid=70&uid=4&uid=3739256&sid=21103766217637

Sukumaran J, Holder MT. 2010. DendroPy: a Python library for phylogenetic computing. Bioinformatics [Internet] 26:1569–1571. Available from: http://www.ncbi.nlm.nih.gov/pubmed/20421198

Wolpert DH, Macready WG. 1997. No Free Lunch Theorems For Optimization. IEEE Trans. Evol. Comput. [Internet] 1:67–82. Available from: http://ti.arc.nasa.gov/m/profile/dhw/papers/78.pdf

Yandell M, Ence D. 2012. A beginner's guide to eukaryotic genome annotation. Nat. Rev. Genet. [Internet] 13:329–342. Available from: http://www.ncbi.nlm.nih.gov/pubmed/22510764

Yang Z. 2007. PAML 4: phylogenetic analysis by maximum likelihood. Mol. Biol. Evol. [Internet] 24:1586–1591. Available from: http://www.ncbi.nlm.nih.gov/pubmed/17483113

Yu C, Zavaljevski N, Desai V, Reifman J. 2011. QuartetS: a fast and accurate algorithm for large-scale orthology detection. Nucleic Acids Res. [Internet] 39:e88. Available from: http://www.pubmedcentral.nih.gov/articlerender.fcgi?artid=3141274&tool=pmcentrez&rendertype=abstract